\documentclass[12pt]{article}

\usepackage{amssymb,amsmath,amsfonts,amsthm,graphicx}

\def\be{\begin{equation}}
\def\ee{\end{equation}}
\def\bea{\begin{eqnarray}}
\def\eea{\end{eqnarray}}

\def\e{\mathbf{e}}
\def\sp{\sigma_+}
\def\udot{\dot{u}}
\def\ex{e_1{}^1}
\def\ey{e_2{}^2}
\def\ez{e_3{}^3}
\def\R{{}^3\!R}
\def\S{{}^3\!S_+}

\def\y{\vartheta}
\def\z{\varphi}

\begin{document}

\title{Spherically Symmetric Cosmology: Resource Paper}

\author{A. A. Coley, W. C. Lim and G. Leon}

\maketitle

\noindent Department of Mathematics and Statistics,
Dalhousie University, Halifax, Nova Scotia, Canada  B3H 3J5 and
Universidad Central de Las Villas, Santa Clara, CP 54830, Cuba.
\\E-mail: aac@mathstat.dal.ca, wlim@princeton.edu, genly@uclv.edu.cu

\begin{abstract}

We use the 1+3 frame formalism to write down the
evolution equations for spherically symmetric models as a
well-posed system of first order PDEs in two variables, suitable
for numerical and qualitative analysis.

\end{abstract}

\section{Introduction}

We shall use the 1+3 frame formalism \cite{EU,WE} to write down the
evolution equations for spherically symmetric models as a
well-posed system of first order PDEs in 2 variables.
The formalism is particularly well-suited for studying
perfect fluid spherically symmetric
models \cite{SSSS}, and especially
for numerical and qualitative analysis, and is useful in
various applications, such as
structure formation in the spherically symmetric dust
Lema\^{\i}tre-Tolman-Bondi model. This preprint is intended as a
resource paper for researchers working in this field.

\section{Spherically symmetric models}

The metric is:%
\footnote{We use $x$ instead of $r$ because $r$ is used to denote the spatial derivative of $H$.}
\be
\label{metric}
        ds^2 = - N^2 dt^2 + (\ex)^{-2} dx^2
        + (\ey)^{-2} (d\y^2 + \sin^2 \y\, d\z^2).
\ee
The Killing vector fields (KVF) are given by \cite{kramer}:
\be
    \partial_\z,\quad
    \cos \z \ \partial_\y - \sin \z \cot \y \ \partial_\z,\quad
        \sin \z \ \partial_\y + \cos \z \cot \y \ \partial_\z.
\ee
The frame vectors in coordinate form are:
\be
    \e_0 = N^{-1} \partial_t
    ,\quad
    \e_1 = \ex \partial_x
        ,\quad
        \e_2 = \ey \partial_\y
        ,\quad
        \e_3 = \ez \partial_\z,
\ee
where $\ez = \ey / \sin \y$. $N$, $\ex$ and $\ey$ are functions of $t$
and $x$.%
\footnote{Note that the frame vectors $\e_2$ and $\e_3$
tangent to the spheres are not group-invariant -- the commutators
$[\e_2, \partial_\z]$ and $[\e_3, \partial_\z]$
are zero, but not with the other two Killing vectors.
The frame vectors $\e_0$ and $\e_1$
orthogonal to the spheres are group-invariant.}

This leads to the following restrictions on the kinematic variables:
\be
    \sigma_{\alpha\beta} = \text{diag}(-2\sp,\sp,\sp),\quad
    \omega_{\alpha\beta} =0,\quad
    \udot_\alpha =(\udot_1,0,0),
\ee
where
\be
    \udot_1 = \e_1 \ln N;
\ee
on the spatial commutation functions:
\be
    a_\alpha = (a_1, a_2, 0),\quad
    n_{\alpha\beta} = \left( \begin{array}{ccc}
            0 & 0 & n_{13}  \\
            0 & 0 & 0   \\
            n_{13} & 0 & 0 \end{array} \right),
\ee
where%
\footnote{The dependence of $a_2$ and $n_{13}$ on $\y$ is due to the fact
that the chosen orthonormal frame is not group-invariant. However, this
is not a concern, since the $\y$ dependence will be hidden.}
\be
    a_1 = \e_1 \ln\ey,\quad
    a_2 = n_{13} = - \frac12 \ey \cot \y;
\ee
and on the matter components:
\be
    q_\alpha = (q_1,0,0),\quad
    \pi_{\alpha\beta} = \text{diag}(-2\pi_+,\pi_+,\pi_+).
\ee
The frame rotation $\Omega_{\alpha\beta}$ is also zero.

Furthermore, $n_{13}$ only appears in the equations together with
$\e_2 n_{13}$ in the form of the Gauss curvature of the spheres
\be
    {}^2\!K := 2(\e_2 - 2 n_{13}) n_{13},
\ee
which simplifies to
\be
    {}^2\!K = (\ey)^2.
\ee
Thus the dependence on $\y$ is hidden in the equations. We will
also use ${}^2\!K$ in place of $\ey$.

The spatial curvatures also simplify to:
\be
        {}^3\!S_{\alpha\beta} = \text{diag}(-2\S,\S,\S),
\ee
with $\R$ and $\S$ given by:
\begin{align}
        \R &= 4 \e_1 a_1 - 6 a_1^2 + 2 {}^2\!K
\\
        \S &= - \tfrac13 \e_1 a_1 + \tfrac13 {}^2\!K.
\end{align}
The Weyl curvature components simplify to:
\be
        E_{\alpha\beta} = \text{diag}(-2E_+,E_+,E_+),\quad
    H_{\alpha\beta} = 0,
\ee
with $E_+$ given by
\be
    E_+ = H\sp + \sp^2 + \S - \tfrac12 \pi_+.
\ee

To simplify notation, we will write
\[
    {}^2\!K,\ \udot_1,\ a_1
\]
as
\[
    K,\ \udot,\ a.
\]
To summarize, the essential variables are
\be
    N, \ex,\ K,\ H,\ \sp,\ a,\ \mu,\ q_1,\ p,\ \pi_+,
\ee
and the auxiliary variables are
\be
    \R,\ \S,\ \udot.
\ee
So far, there are no evolution equations for $N$, $p$ and $\pi_+$, and
they need to be specified by a temporal gauge (for $N$), and by a fluid
model (for $p$ and $\pi_+$).

The evolution equations are now:%
\footnote{We include a non-negative $\Lambda$.}
\begin{align}
    \e_0 \ex &= (-H+2\sp) \ex
\\
    \e_0 K &= -2(H+\sp)K
\\
    \e_0 H &= - H^2 - 2 \sp^2 + \tfrac13 (\e_1 + \udot - 2 a)\udot
        - \tfrac16(\mu+3p) + \tfrac13 \Lambda
\\
    \e_0 \sp &= -3H \sp - \tfrac13(\e_1 + \udot + a)\udot
        - \S + \pi_+
\\
    \e_0 a &= (-H+2\sp) a - (\e_1 + \udot)(H+\sp)
\\
    \e_0 \mu &= -3H(\mu+p) - (\e_1+2\udot-2a)q_1 - 6\sp\pi_+
\\
    \e_0 q_1 &= (-4H+2\sp)q_1 - \e_1 p -(\mu+p)\udot
            + 2(\e_1+\udot-3a)\pi_+.
\end{align}
The constraint equations are the Gauss and Codazzi constraints, and the
definition of $a$:
\begin{align}
    0 &= 3H^2 + \tfrac12 \R - 3 \sp^2 - \mu - \Lambda
\\
    0 &= -2 \e_1(H+\sp) + 6 a \sp + q_1
\\
    0 &= (\e_1 - 2 a) K,
\end{align}
where the spatial curvatures are given by
\begin{align}
    \R &= 4 \e_1 a - 6 a^2 + 2 K
\\
    \S &= - \tfrac13 \e_1 a + \tfrac13 K.
\end{align}

\section{The matter and gauge}

There are various choices for the matter.

\subsection{Perfect fluid}

A perfect fluid is defined by \cite{WE} \be
    T_{ab} = \hat{\mu} u_a u_b + \hat{p} ( g_{ab} + u_a u_b).
\ee
with $\hat{p}$ to be specified.
In general, the 4-velocity vector $\mathbf{u}$ of the perfect fluid is not
aligned with the vector $\e_0$ of a chosen temporal gauge. In spherically
symmetric models, $\mathbf{u}$ is allowed to be of the form
\be
    \mathbf{u} = \Gamma(\e_0 + v \e_1),\quad
    \Gamma = (1-v^2)^{-\frac12}.
\ee
We choose a linear equation of state for the perfect fluid:
\be
\label{linear_eos}
    \hat{p} = (\gamma-1) \hat{\mu},
\ee
where $\gamma$ is a constant satisfying $1 \leq \gamma < 2$.
Then we obtain for the tilted fluid:
\begin{align}
    \mu &= \frac{G_+}{1-v^2} \hat{\mu}
\\
    p &= \frac{(\gamma-1)(1-v^2) + \frac13\gamma v^2}{G_+} \mu
\\
    q_1 &= \frac{\gamma \mu}{G_+} v
\\
    \pi_+ &= - \frac13 \frac{\gamma \mu}{G_+} v^2,
\end{align}
where $G_\pm = 1 \pm (\gamma-1)v^2$.
Thus $p$, $q_1$ and $\pi_+$ are given in terms of $\mu$ and $v$.
These are then substituted into the evolution and constraint equations.

The evolution equations for $\mu$ and $q_1$ now give (in terms of $\mu$
and $v$)
\begin{align}
    \e_0 \mu &= - \frac{\gamma v}{G_+} \e_1 \mu
        - \frac{\gamma G_-}{G_+^2} \mu \e_1 v
        - \frac{\gamma}{G_+} \mu \left[
            (3+v^2)H + 2v(\udot-a) - 2 v^2 \sp \right]
\\
\label{dv}
    \e_0 v &= - \frac{(\gamma-1)(1-v^2)^2}{\gamma G_- \mu} \e_1 \mu
    + \frac{[(3\gamma-4)-(\gamma-1)(4-\gamma)v^2]v}{G_+ G_-} \e_1 v
\notag\\
    &\qquad
        - \frac{(1-v^2)}{G_-} \left[
        -(3\gamma-4)v H - 2 v \sp + G_- \udot +2(\gamma-1)v^2 a
                    \right].
\end{align}

\subsection{Scalar fields and anisotropic fluid}

The total energy-momentum tensor of a non-interacting scalar field
$\phi$ with a self-interaction potential $V(\phi)$  is \be T^{sf}_{a
b}=\phi_{;a}\phi_{;b}-g_{a
b}(\tfrac12\phi_{;c}\phi^{;c}+V(\phi)),\ee where $\phi=\phi(t,x).$
In particular, exponential potentials have been the subject of much interest and
arise naturally from theories of gravity such as scalar-tensor
theories or string theory \cite{exponential}. Spherically symmetric scalar field
models have been studied in
\cite{Genly}.

A spherically symmetric model can also admit an anisotropic fluid matter source, 
in which the energy momentum tensor
has energy density $\mu$, a pressure $p_{||}$ parallel to the
radial unit normal and a perpendicular pressure $p_\perp$.
Fluids with an anisotropic pressure have been studied in the
cosmological context for a number of reasons \cite{Kras:1996}. An
energy-momentum tensor of this form formally arises if the source
consists of two perfect fluids with distinct four-velocities, a
heat conducting viscous fluid and a perfect fluid and a magnetic field;
in addition, a cosmic
string and a global monopole are of the form of an anisotropic fluid.
Most importantly, perhaps, a contribution in the form of an
anisotropic fluid arises when averaging the Einstein equation to
obtain the averaged field equations in spherically symmetric
geometries \cite{ColPel}.

\subsection{Temporal gauge}

The common temporal gauges used in spherically symmetric cosmological models
are the {\em synchronous gauge} and the {\em separable
area gauge}.  The synchronous gauge is useful when used with a dust perfect
fluid ($\gamma=1$) because the dust perfect fluid has zero acceleration
($\udot=0$). This gives the Lema\^{\i}tre-Tolman-Bondi models.
It may also be useful when used with a non-dust tilted perfect
fluid. The un-normalized system is well-posed
when the Gauss and Codazzi constraints are used to eliminate the spatial
derivatives. $H$-normalization preserves well-posedness.

    The separable area gauge has a special case ($a=0$), called the
{\em timelike area gauge}. The un-normalized system is well-posed when
$\e_0(H+\sp)$ is used, and the Gauss constraint is solved for $H-\sp$.
$(H+\sp)$-normalization preserves well-posedness, while $H$-normalization
does not.

\section{Special cases with extra Killing vectors}

Spherically symmetric models with more than 3 KVF are either
spatially homogeneous or static. Let us  discuss the spatially
homogeneous cosmological models.
Spatially homogeneous spherically symmetric models consist of two
disjoint sets of models: the Kantowski-Sachs models and the 
Friedmann-Lema\^{\i}tre-Robertson-Walker (FLRW) models.
Static and self-similar spherically symmetric models have been studied in
\cite{Genly,static,SSSS}.

\subsection{The Kantowski-Sachs models}

The spatially homogeneous spherically symmetric models (that has 4
Killing vectors, the fourth being $\partial_x$) are the so-called
Kantowski-Sachs models \cite{kramer}.
The metric (\ref{metric}) simplifies to
\be
        ds^2 = - N(t)^2 dt^2 + (\ex(t))^{-2} dx^2
                + (\ey(t))^{-2} (d\y^2 + \sin^2 \y\, d\z^2);
\ee
i.e., $N$, $\ex$ and $\ey$ are now independent of $x$.

The spatial derivative terms $\e_1(\ )$ vanish and as a
result $a=0=\udot$. Since $\udot=0$, the temporal gauge is
synchronous and we can set $N$ to any positive function of $t$.

The Codazzi constraint restricts the source by
\be
    q_1 = 0.
\ee
$p$ and $\pi_+$ are still unspecified.

The evolution equations for Kantowski-Sachs models with unspecified
source are:
\begin{align}
        \e_0 \ex &= (-H+2\sp) \ex
\\
        \e_0 K &= -2(H+\sp)K
\\
        \e_0 H &= - H^2 - 2 \sp^2
                - \tfrac16(\mu+3p) + \tfrac13 \Lambda
\\
        \e_0 \sp &= -3H \sp - \tfrac13 K + \pi_+
\\
        \e_0 \mu &= -3H(\mu+p) - 6\sp\pi_+
\end{align}
The remaining constraint equation is the Gauss constraint:
\be
        0 = 3H^2 + K - 3 \sp^2 - \mu - \Lambda.
\ee
The spatial curvatures are given by
\begin{align}
        \R &= 2 K
\\
        \S &= \tfrac13 K.
\end{align}

\subsection{The FLRW models}

Spatially homogeneous spherically symmetric models, that are not
Kantowski-Sachs, are the
Friedmann-Lema\^{\i}tre-Robertson-Walker (FLRW) models (with or without
$\Lambda$).
The source must be of the form of a comoving perfect fluid (or vacuum).

The metric has the form
\be
        ds^2 = - N(t)^2 dt^2 + \ell^2(t) dx^2
                + \ell^2(t) f^2(x) (d\y^2 + \sin^2 \y\, d\z^2),
\ee
with
\be
\label{fx_FLRW}
    f(x) = \sin x,\ x,\ \sinh x,
\ee
for closed, flat, and open FLRW models respectively.
The frame coefficients are given by $\ex = \ell^{-1}(t)$ and $\ey =
\ell^{-1}(t) f^{-1}(x)$. Then $\sp = \frac13\e_0 \ln(\ex/\ey)$
vanishes.
$N=N(t)$ implies that $\udot=0$; i.e.,
the temporal gauge is synchronous, and we can set $N$ to any positive
function of $t$.
The Hubble scalar $H = \e_0 \ln \ell(t)$ is also a function of
$t$.
{\footnote {We shall not list the KVFs as they are complicated in spherically
symmetric coordinates and not needed here.}}

For the spatial curvatures, $\S$ does vanish because (\ref{fx_FLRW})
implies $\e_1 a = K$,%
\footnote{That $\e_1 a$ does not vanish is consistent with
 the frame vector $\e_1$ not being group-invariant.}
 while $\R$ simplifies to
\be
    \R = \frac{6k}{\ell^2},\quad
    k = 1,0,-1,
\ee
for closed, flat, and open FLRW respectively.

The evolution equation for $\sp$ and the Codazzi constraint then imply
that $\pi_+ =0= q_1$; i.e., the source is a comoving perfect fluid, with
unspecified pressure $p$.

The evolution equations simplify to:
\begin{align}
    \e_0 \ell &= H \ell
\\
    \e_0 H    &= - H^2 - \tfrac16(\mu+3p) + \tfrac13 \Lambda
\\
    \e_0 \mu  &= -3H(\mu+p)
\end{align}
The Gauss constraint simplifies to
\be
    0 = 3H^2 + \frac{3k}{\ell^2} - \mu - \Lambda,\quad k=1,0,-1.
\ee
Note that $\mu$ and $p$ also depend on $t$ only, and that $p$ is not
specified yet.

The vacuum cases are the de Sitter model ($\Lambda>0$, $k=0$),
the model with $\Lambda>0$, $k=1$,
the model with $\Lambda>0$, $k=-1$,
the Milne model ($\Lambda=0$, $k=-1$),
and the Minkowski spacetime ($\Lambda=0$, $k=0$), which is also static.
The model with $\Lambda>0$, $k=1$ is past asymptotic to the de Sitter
model with negative $H$ and is future asympotic to the de Sitter model
with positive $H$.
The model with $\Lambda>0$, $k=-1$ (and positive $H$) is past asymptotic
to the Milne model and is future asympotic to the de Sitter model with
positive $H$.

\section{Synchronous gauge, tilted perfect fluid}

We shall investigate perfect fluid models with linear equation of state
using the synchronous gauge.
We shall simplify the equations step-by-step, by choosing the synchronous
gauge, eliminating spatial derivatives, and specifying the perfect fluid.

The equations in synchronous gauge ($\udot=0$) are:
\begin{align}
        \e_0 \ex &= (-H+2\sp) \ex
\\
        \e_0 K &= -2(H+\sp)K
\\
        \e_0 H &= - H^2 - 2 \sp^2 - \tfrac16(\mu+3p) + \tfrac13 \Lambda
\\
\label{dsp_synch}
        \e_0 \sp &= -3H \sp - \S + \pi_+
\\
\label{da_synch}
        \e_0 a &= (-H+2\sp) a - \e_1(H+\sp)
\\
        \e_0 \mu &= -3H(\mu+p) - (\e_1-2a)q_1 - 6\sp\pi_+
\\
        \e_0 q_1 &= (-4H+2\sp)q_1 - \e_1 p + 2(\e_1-3a)\pi_+.
\end{align}
The constraint equations are:
\begin{align}
\label{CG_synch}
        0 &= 3H^2 + \tfrac12 \R - 3 \sp^2 - \mu - \Lambda
\\
\label{CC_synch}
        0 &= -2 \e_1(H+\sp) + 6 a \sp + q_1
\\
        0 &= (\e_1 - 2 a) K.
\end{align}
where the spatial curvatures are given by
\begin{align}
        \R &= 4 \e_1 a - 6 a^2 + 2 K
\\
        \S &= - \tfrac13 \e_1 a + \tfrac13 K.
\end{align}

The evolution equations (\ref{dsp_synch}) and (\ref{da_synch}) contain
spatial derivative terms, but these can be replaced using the
constraints (\ref{CG_synch}) and (\ref{CC_synch}):
\begin{align}
    \e_1 a &= - \tfrac32 H^2 + \tfrac32 a^2 - \tfrac12 K
        + \tfrac32 \sp^2 + \tfrac12 \mu + \tfrac12 \Lambda
\\
    \e_1 (H+\sp) &= 3a\sp + \tfrac12 q_1.
\end{align}
As a result, equations (\ref{dsp_synch}) and (\ref{da_synch}) now read:
\begin{align}
        \e_0 \sp &= -3H \sp - \tfrac12 H^2 + \tfrac12 a^2 - \tfrac12 K
                + \tfrac12 \sp^2 + \tfrac16 \mu + \tfrac16 \Lambda
        + \pi_+
\\
        \e_0 a &= -(H+\sp) a - \tfrac12 q_1.
\end{align}
The benefit here is that the evolution equations for the geometric part
are now free of spatial derivative terms.

The spatial curvatures are given by
\begin{align}
        \R &= - 6 H^2 + 6 \sp^2 + 2 \mu + 2 \Lambda
\\
        \S &= \tfrac12 H^2 - \tfrac12 a^2 + \tfrac12 K
                - \tfrac12 \sp^2 - \tfrac16 \mu - \tfrac16 \Lambda.
\end{align}

Lastly, we specify the perfect fluid with linear equation of state. From
equations (\ref{linear_eos})--(\ref{dv}) and the above equations, the
final form of the system is:
\begin{align}
\label{dex_pfs}
        \e_0 \ex &= (-H+2\sp) \ex
\\
        \e_0 K &= -2(H+\sp)K
\\
        \e_0 H &= - H^2 - 2 \sp^2
        - \frac{(3\gamma-2+(2-\gamma)v^2)}{6G_+} \mu
        + \tfrac13 \Lambda
\\
        \e_0 \sp &= -3H \sp - \tfrac12 H^2 + \tfrac12 a^2 - \tfrac12 K
                + \tfrac12 \sp^2
                + \frac{(1-(\gamma+1)v^2)}{6G_+} \mu
        + \tfrac16 \Lambda
\\
        \e_0 a &= -(H+\sp) a - \frac{\gamma v}{2G_+} \mu
\\
        \e_0 \mu &+ \frac{\gamma v}{G_+} \e_1 \mu
                + \frac{\gamma G_-}{G_+^2} \mu \e_1 v
\notag\\
    &=
                - \frac{\gamma}{G_+} \mu \left[
                        (3+v^2)H - 2va - 2 v^2 \sp \right]
\\
\label{dv_pfs}
        \e_0 v &+ \frac{(\gamma-1)(1-v^2)^2}{\gamma G_- \mu} \e_1 \mu
        - \frac{[(3\gamma-4)-(\gamma-1)(4-\gamma)v^2]v}{G_+ G_-} \e_1 v
\notag\\
        &=
                - \frac{(1-v^2)}{G_-} \left[
                -(3\gamma-4)H - 2 \sp +2(\gamma-1)v a
                                        \right] v,
\end{align}
where $G_\pm = 1 \pm (\gamma-1)v^2$.
The constraints are:
\begin{align}
        \e_1 a &= - \tfrac32 H^2 + \tfrac32 a^2 - \tfrac12 K
                + \tfrac32 \sp^2 + \tfrac12 \mu + \tfrac12 \Lambda
\\
        \e_1 (H+\sp) &= 3a\sp + \frac{\gamma v}{2G_+} \mu
\\
        \e_1 K &= 2 a K.
\end{align}
The spatial curvatures are given by
\begin{align}
        \R &= - 6 H^2 + 6 \sp^2 + 2 \mu + 2 \Lambda
\\
        \S &= \tfrac12 H^2 - \tfrac12 a^2 + \tfrac12 K
                - \tfrac12 \sp^2 - \tfrac16 \mu - \tfrac16 \Lambda.
\end{align}

\subsection{Well-posedness}

We now show that the system is well-posed for $\gamma \geq 1$.

The coefficient matrix for the spatial derivative terms is:
\footnote{Strictly speaking, we should also include the factor $\ex/N$ in the
matrix, but the result on well-posedness is the same.}
\be
    \left(
\begin{array}{cc}
    \dfrac{\gamma v}{G_+} & \dfrac{\gamma G_-}{G_+^2} \mu \\
    \dfrac{(\gamma-1)(1-v^2)^2}{\gamma G_- \mu} &
        - \dfrac{[(3\gamma-4)-(\gamma-1)(4-\gamma)v^2]v}{G_+ G_-}
\end{array}
    \right).
\ee
Its eigenvalues are
\be
    \frac{(2-\gamma)v \pm \sqrt{\gamma-1}(1-v^2)}{G_-},
\ee
with corresponding eigenvectors (for example)
\be
    \left(
\begin{array}{c}
    1 \\
    \dfrac{\sqrt{\gamma-1}(1-v^2)G_+}{\mu\gamma(1+\sqrt{\gamma-1}v)^2}
\end{array}
    \right)
    ,\quad
        \left(
\begin{array}{c}
        1 \\
        -\dfrac{\sqrt{\gamma-1}(1-v^2)G_+}{\mu\gamma(1-\sqrt{\gamma-1}v)^2}
\end{array}
        \right).
\ee

The matrix is diagonalizable for $\gamma > 1$, with
$c_s = \sqrt{\gamma-1}$ being the speed of sound in the perfect fluid.
The system (\ref{dex_pfs})--(\ref{dv_pfs}) is thus well-posed for $\gamma
> 1$. For $\gamma < 1$ the system is elliptic and not well-posed.

\section{Irrotational dust (Lema\^{\i}tre-Tolman-Bondi model)}

The Lema\^{\i}tre-Tolman-Bondi (LTB) model
\cite{lemaitre,Kras:1996} is the spherically symmetric dust
solution of the Einstein equations which can be regarded as a
generalization of the FLRW universe. LTB metrics with dust source
and a comoving and geodesic 4-velocity constitute a well known
class of exact solutions of Einstein's field equations
\cite{kramer, Kras:1996}.

For the dust case $\gamma=1$ with zero vorticity, we can use the
freedom within the synchronous gauge to set $v=0$, so that the
synchronous frame is comoving with the perfect fluid and we obtain:
\begin{align}
        \e_0 \ex &= (-H+2\sp) \ex
\\
        \e_0 K &= -2(H+\sp)K
\\
        \e_0 H &= - H^2 - 2 \sp^2
                - \tfrac16 \mu
                + \tfrac13 \Lambda
\\
        \e_0 \sp &= -3H \sp - \tfrac12 H^2 + \tfrac12 a^2 - \tfrac12 K
                + \tfrac12 \sp^2
                + \tfrac16 \mu
                + \tfrac16 \Lambda
\\
        \e_0 a &= -(H+\sp) a
\\
        \e_0 \mu &= - 3H\mu.
\end{align}
Notice that the system is completely free of spatial derivatives, and is
thus well-posed.

The constraints are:
\begin{align}
        \e_1 a &= - \tfrac32 H^2 + \tfrac32 a^2 - \tfrac12 K
                + \tfrac32 \sp^2 + \tfrac12 \mu + \tfrac12 \Lambda
\\
        \e_1 (H+\sp) &= 3a\sp
\\
        \e_1 K &= 2 a K.
\end{align}
The spatial curvatures are given by
\begin{align}
        \R &= - 6 H^2 + 6 \sp^2 + 2 \mu + 2 \Lambda
\\
        \S &= \tfrac12 H^2 - \tfrac12 a^2 + \tfrac12 K
                - \tfrac12 \sp^2 - \tfrac16 \mu - \tfrac16 \Lambda.
\end{align}
Further simplifications with $a$ and $K$ are possible.

A suitable normalization factor is $\beta=H+\sp$,
introduced for $G_2$ models in \cite{WE,EU}, and used for the LTB model in \cite{Sussman}.
With this normalization, it can be shown that at late times
the LTB solutions that are ever-expanding will tend to the isotropic and homogeneous Milne solution, with the following rates:
\begin{gather}
 \beta \sim e^{-\tau},\qquad \frac{a^2-K}{\beta^2} - 1 \sim e^{-\tau},
\\
 \frac{\sp}{\beta} \sim \tau e^{-\tau},\qquad \frac{\mu}{3\beta^2} \sim e^{-\tau}.
\end{gather}
That is, the rates are the same for all dust observers,
although the multiplicative ``constants" depend on the radius.
This dependency reveals itself in the leading order of ratios of variables such as the density contrast.

\subsection{Structure formation}

Structure formation in the LTB model has been studied in \cite{KH}.
More recently, the LTB inhomogeneous dust solutions have been  examined
numerically and qualitatively as a 3--dimensional dynamical
system, in terms of an average
density parameter, $\langle\Omega\rangle$ (which behaves dynamically like 
the usual
$\Omega$ in FLRW dust spacetimes), and a
shear parameter and a density contrast function which convey the
effects of inhomogeneities \cite{Sussman}. The evolution equations for the
averaged variables are formally identical to those of an
equivalent FLRW cosmology, and are an alternative set of evolution equations
to those presented above. In particular, the phase space evolution of
structure formation scenario was examined in \cite{Sussman}.

\end{document}